# Variable Neighborhood Search for the Bin Packing Problem with Compatible Categories

Moura Santos, Luiz F. O.[1], Yoshizaki, Hugo T.Y.[1], and Cunha, Claudio B.[2]

1: Department of Production Engineering, University of São Paulo, São Paulo, SP, Brazil
2: Department of Transportation Engineering, University of São Paulo, São Paulo, SP, Brazil

**Abstract**

Bin Packing with Conflicts (BPC) are problems in which items with compatibility constraints must be packed in the least number of bins, not exceeding the capacity of the bins and ensuring that non-conflicting items are packed in each bin. In this work, we introduce the Bin Packing Problem with Compatible Categories (BPCC), a variant of the BPC in which items belong to conflicting or compatible categories, in opposition to the item-by-item incompatibility found in previous literature. It is a common problem in the context of last mile distribution to nanostores located in densely populated areas. To efficiently solve real-life sized instances of the problem, we propose a Variable Neighborhood Search (VNS) metaheuristic algorithm. Computational experiments suggest that the algorithm yields good solutions in very short times while compared to linear integer programming running on a high-performance computing environment.

## 1. Introduction

Last-mile delivery in developing countries remains a significant challenge to retailers and shippers, as they commonly distribute to several tiny stores located in densely populated areas, named nanostores (Fransoo, Blanco, & Mejia-Argueta, 2017). These stores usually keep small amounts of a limited set of stock keeping units (SKUs), forcing companies to make frequent deliveries using small vehicles packed with a large variety of SKUs from different suppliers.

In this paper, we tackle the last-mile delivery to nanostores. The problem is modeled as a Bin Packing Problem with Compatible Categories (BPCC), a variant of the classical Bin Packing Problem (BPP) (Delorme, Iori, & Martello, 2016) and the Bin Packing with Conflicts (BPC) (Gendreau, Laporte, & Semet, 2004). In the BPCC, items refer to products that belong to different categories, such as cleaning and hygiene products, groceries, dairy products, canned foods, alcohol and so on. In this way, due to contamination and risk issues, each pair of categories may or may not be compatible, leading incompatible items not to be packed in the same bin (vehicle).

Despite the literature on the BPP being extensive, works focused on bin packing with conflicts (BPC) are still scarce. In fact, to our knowledge, this is the first appearance of the BPC in which conflicts are between categories of products, in opposition to the item-by-item incompatibility addressed in previous literature.

Also, it is known that the BPP is NP-hard (Delorme et al., 2016), what makes it challenging to solve large instances of the problem that are commonly found in practice using optimization solely. In this way, we propose a Variable Neighborhood Search (VNS) algorithm to solve it and test it against a set of instances derived from difficult bin packing benchmark problems presented by Delorme, Iori, & Martello (2016). Thus, the objectives of this paper are: (i) present a new mathematical formulation for the BPCC; and (ii) propose a VNS heuristic to solve it efficiently.

The remaining of this article is organized as follows: a brief literature review is presented in Section 2; the mathematical formulation for the BPCC is given in Section 3; in Section 4 we propose a VNS algorithm to solve the problem; Section 5 contains the computational experiments, results, and discussions; and finally, conclusions are presented in Section 6.

## 2. Literature review

The last-mile delivery problem faced by traditional retail suppliers and shippers has been modeled as a combination of loading and routing problems by the literature, such as the 3-dimensional Loading Capacitated Vehicle Routing Problem (3L-CVRP) introduced by Gendreau, Iori, Laporte, & Martello (2006). The loading part of the 3L-CVRP provides similar solutions as the Bin Packing Problem (BPP) formulation, where the loaded good's weight cannot exceed the vehicle's weight capacity. However, compatibility constraints between items are not addressed in neither of these problems.

Conflicts between individual pairs of objects have been introduced to the BPP by Jansen & Öhring (1997). The authors propose and evaluate different approximation algorithms for solving the BPC, relying on a conflict graph. Such representation of the problem has also been used by other authors, such as Gendreau, Laporte, & Semet (2004) and Fernandes-Muritiba, Iori, Malaguti, & Toth (2010), who explicitly address the problem of Bin Packing with Conflicts (BPC). However, these works only consider compatibility among items in a pairwise manner, i.e., items are not divided into categories as in our BPCC formulation, which addresses this gap.

Also, many different metaheuristics have been used to solve the BPP and its variants. The use of Variable Neighborhood Search (VNS), for instance, have been reported by Fleszar & Hindi (2002) and Hemmelmayr, Schmid, & Blum (2012). Another example is Loh, Golden, & Wasil (2008), which presented a weight annealing heuristic to obtain high-quality solutions very quickly. Stawowy (2008), on the other hand, used an evolutionary-based algorithm, similar to an Iterated Local Search approach.

Usually, these metaheuristics start with initial solutions obtained using greedy heuristics, such as First-Fit and the Best-Fit algorithms, or First-Fit Decreasing (FFD) and Best-Fit Decreasing (BFD) if items are initially sorted in non-increasing order of weights. The interested reader may refer to Delorme et al. (2016) for a summary of metaheuristics and greedy heuristics applied to BPPs and its variants.

## 3. Bin Packing with Compatible Categories (BPCC)

The BPCC is defined given a set $C = \{1,2,...,p\}$ of categories; a set $J = \{1,2,3,...,n\}$ of items, each item $j \in J$ weighting $w_j \in \mathbb{Z}^+$ and belonging to a category $cat_j \in C$; an infinite number of identical bins of capacity $b \in \mathbb{Z}^+$; and a compatibility matrix $C_{pxp}$ of binary elements $c^{kl}$ that equals 1 if the category $k \in C$ is compatible with category $l \in C$ and 0 otherwise. Items belonging to categories that are not compatible cannot be assigned to the same bin. We also define $p$ subsets $J^k \subset J$ that comprises all items $j \in J$ such that $cat_j = k, \forall j \in J^k$ and assume $b \geq w_j \ \forall j \in J$. The BPCC aims to use the minimum number of bins while assigning all items to the bins, making sure that each bin does not exceeds its weight capacity and only contains items from compatible categories.

The BPCC is modeled using three sets of binary decision variables: (i) $y_i$ that equals 1 if bin $i$ is used and 0 otherwise; (ii) $x_{ij}$ that equals 1 if item $j$ is assigned to bin $i$ and 0 otherwise; and (iii) $f_i^k$ that equals 1 if at least one item belonging to category $k$ is assigned to bin $i$ and 0 otherwise. Thus, the linear integer programming model for the problem is given as follows:

$$min \ z = \sum_{i=1}^{n} y_i \qquad (1)$$

Subject to:

$$\sum_{j \in J} x_{ij} * w_j \leq by_i \qquad i = 1,2,...,n \qquad (2)$$

$$\sum_{i=1}^{n} x_{ij} = 1 \qquad j \in J \qquad (3)$$

$$x_{ij} \leq f_i^k \qquad \begin{matrix} i \in \{1,2,...,n\} \\ k \in C, \ j \in J^k \end{matrix} \qquad (4)$$

$$f_i^k + f_i^l \leq 1 \qquad \begin{array}{l} i = 1,2,\ldots,n \\ \forall k,l \in C: c^{kl} = 0 \end{array} \qquad (5)$$

$$y_i \geq y_s \qquad i,s \in \{1,2,\ldots,n\}: i < s \qquad (6)$$

$$y_i, x_{ij}, f_i^k \in \{0,1\} \qquad \begin{array}{l} \forall i,j \in J \\ \forall k \in C \end{array} \qquad (7)$$

The objective function (1) minimizes the total number of required bins. Constraints (2) are the capacity constraints, while constraints (3) ensure that each item is assigned to precisely one bin. Constraints (4) link variables $x_{ij}$ and $f_i^k$, such that if an item belonging to category $k$ is assigned to a bin $i$, $f_i^k$ equals 1; $M$ denotes a sufficiently large number, e.g. the number of items $n$. Constraints (5) enforce that for every pair of incompatible categories, at most one of them is present in a bin, for every used bin. Constraints (6) are not mandatory, but they set an indexing of the used bins that aims to help finding optimal solutions more easily.

## 4. Variable Neighborhood Search for the BPCC

The variable neighborhood search (VNS) is a simple and efficient metaheuristic that explores the solution space using systematic changes of neighborhoods within local search procedures (Hansen & Mladenović, 2001). It is given an initial solution $x_0$ and a predetermined set of neighborhood structures $\mathcal{N}_k = (k_1, k_2, \ldots, k_{max})$, being $\mathcal{N}_k(x)$ the set of solutions in the $k$-th neighborhood of the incumbent solution $x$. In our case, we use $k_{max} = 4$. The VNS updates the current solution only if improvement has been made in that iteration.

An essential factor to any neighborhood search procedure is the choice of the objective function (Fleszar & Hindi, 2002). In a BPP that minimizes the number of bins, many solutions have the same objective value, since items can be arranged in several different ways. So, we use a fitness function $h(x)$ to differentiate solutions, based on the assumption that better solutions tend to have fuller bins. It is given by Expression (8), where $nb$ is the number of used bins in the current solution and assuming $N_i$ as the set of all items $j \in J$ that are assigned to bin $i$ (i.e., such that $x_{ij} = 1$), In addition to $h(x)$, the original objective function (1) is also considered when updating the best solution $x_{best}$, meaning that only solutions with the same or fewer total number of bins are considered.

$$h(x) = \sum_{i=1}^{nb} \sum_{j \in N_i} \left(\frac{w_j x_{ij}}{b}\right)^2 \qquad (8)$$

In our VNS, we use an adaptive method $ChooseNeighbourhood(s_1, s_2, s_3, s_4)$ that selects the neighborhood structure to be used in the current iteration, according to a probability based on scores $s_1$, $s_2$, $s_3$ and $s_4$, and computed from the number of successful iterations using each shaking operator, like Hemmelmayr et al. (2012). The pseudocode of our VNS heuristic is presented in Algorithm 1.

Shaking procedures introduce random perturbations to the incumbent solution, to find better ones in far neighborhoods. In this way, function $Shaking(x, k)$ randomly generates a $x'$ solution in the $k$-th neighborhood of $x$, i.e. $\mathcal{N}_k(x)$. We propose a set of four neighborhood operators:

- $\mathcal{N}_1$ randomly chooses a category $k \in C$ for which all items are removed from its bins. These items are then repacked using a greedy FFD-based algorithm for each category (named FFCD);
- $\mathcal{N}_2$ is like $\mathcal{N}_1$, but randomly chooses two (instead of one) categories to be removed from the bins and then repacked;
- $\mathcal{N}_3$ has two random steps. First, it is drawn a number of bins to be fully emptied, respecting a parameter $\alpha$ that sets the maximum percentage of bins allowed to be erased within a solution. Bins to be emptied are then randomly chosen among all bins. Lastly, FFCD is used to repack the removed items; and
- $\mathcal{N}_4$ works in the same way as $\mathcal{N}_3$, but explores farther away neighborhoods by emptying a larger number of bins than $\mathcal{N}_3$. It is also driven by a parameter $\beta > \alpha$ that sets the maximum percentage of bins allowed to be erased within the solution.

```
Start: VNS for the BPCC.
 1:   $x \leftarrow x_0$;
 2:   $x_{best} \leftarrow x$;
 3:   $s_1 \leftarrow 1; s_2 \leftarrow 1; s_3 \leftarrow 1; s_4 \leftarrow 1$;
 4:   While (stopping criteria not met) do
 5:        $k \leftarrow ChooseNeighbourhood(s_1, s_2, s_3, s_4)$;
 6:        $x' \leftarrow Shaking(x, k)$;
 7:        $x'' \leftarrow LocalSearches(x')$;
 8:        If $\left( h(x'') > h(x_{best}) \text{ and } z(x'') \leq z(x_{best}) \right)$ then
 9:             $x_{best} \leftarrow x''$;
10:             $s_k \leftarrow s_k + 1$;
11:        End − if;
12:        $x \leftarrow x_{best}$;
13: End − while;
End: VNS for the BPCC.
```

Algorithm 1. VNS for the BPCC

After a shaking using $\mathcal{N}_k$, $LocalSearches(x')$ perform two local search procedures every iteration:

- $\mathcal{L}_1$ aims to increase the average density of the bins, based on moving items from one bin to another. The procedure searches every bin with less used capacity than a parameter $\gamma$. It then checks if any item fits best into another bin, moving it if so. However, moves are performed only if they lead to a better capacity usage, by moving items from bins with lower used capacities to other more occupied. Also, a move is performed only if it does not lead to a violation in terms of the bin capacity or generate a category conflict.
- $\mathcal{L}_2$, on the other hand, is based on swaps of two items assigned to different bins. It first ranks all possible swaps based on their improvements to the fitness function (expression 8), only listing swaps with positive improvements. These improvements are incrementally calculated as described by Fleszar & Hindi (2002), calculating only the changes to the fitness function due to each swap. $\mathcal{L}_2$ ends after all possible swaps from the rank have been made.

The VNS ends whenever one of three stopping criteria is met: a number $\lambda$ of iterations without improvements is reached; a total of $\varphi$ iterations is reached; or an optimal solution is found (i.e., its objective equals the continuous lower bound $L_{CONT} = \left\lceil \frac{\sum_{i=1}^{n} w_i}{b} \right\rceil$).

## 5. Computational Experiments

To assess the performance of the proposed VNS heuristic, in our experiments we used BPCC instances derived from the *augmented non*-IRUP (ANI) and the *augmented* IRUP (AI) sets of difficult instances proposed by Delorme, Iori, & Martello (2016). The instances comprise two sets of 50 ANI and 50 AI instances. Each ANI set has $n \in \{201, 402\}$ items, while the AI sets have $n \in \{202, 403\}$ items. Also, as the BPCC is related to last-mile distribution to nanostores, in which vehicles don't usually travel at full capacity, each instance was tested considering four capacity factors: 100%, 120%, 150% and 200% of its original capacity, totalling 800 tested instances.

The VNS was coded in C++ and ran on an Intel® Core™ i5-4690 CPU @ 3.5GHz with 16 GB of RAM personal computer (PC). We used the following parameters: $\lambda = 200$; $\varphi = 2000$; $\alpha = 0.25$; $\beta = 0.5$; and $\gamma = 1$. Also, as there were lots of instances to test and NP-hard optimization using the mathematical formulation given by expressions (1) to (7) takes very long times to run in the PC, we used a high-performance computing (HPC) resource. The HPC is a cluster with 64 physical servers (nodes); each server has an Intel® Xeon® CPU E7-2870 @ 2.40 GHz with 10 cores, 20 threads, and 512 GB of RAM, and runs the solver IBM ILOG CPLEX v12.6. However, only one node was used for the experiments. The compatibility matrix of categories is presented in Table 1, while results are presented in Table 2.

|  | $C_1$ | $C_2$ | $C_3$ | $C_4$ | $C_5$ | $C_6$ |
|---|---|---|---|---|---|---|
| $C_1$ | 1 | 0 | 1 | 0 | 0 | 0 |
| $C_2$ | 0 | 1 | 0 | 1 | 1 | 1 |
| $C_3$ | 1 | 0 | 1 | 1 | 0 | 0 |
| $C_4$ | 0 | 1 | 1 | 1 | 1 | 1 |
| $C_5$ | 0 | 1 | 0 | 1 | 1 | 0 |
| $C_6$ | 0 | 1 | 0 | 1 | 0 | 1 |

Table 1. Compatibility matrix

| Instance Type | # Items | Cap. Fact (%) | CPLEX 12.6 (HPC) | | | | VNS (PC) | | | |
|---|---|---|---|---|---|---|---|---|---|---|
| | | | # Opt.[1] | Total # Bins | Mean Time[3] (s) | Std. Dev. Time[3] (s) | # Opt.[2] | Total # Bins (Best) | Mean Time[4] (s) | Std. Dev. Time[4] (s) |
| AI | 202 | 100 | 5 | 3300 | 626.898 | 931.188 | 5 | 3300 | 6.637 | 0.282 |
| | | 120 | 49 | 2771 | 296.954 | 1047.426 | 50 | 2771 | 0.182 | 0.228 |
| | | 150 | 50 | 2200 | 5.059 | 1.774 | 50 | 2200 | *N/A* | *N/A* |
| | | 200 | 50 | 1650 | 6.171 | 2.845 | 50 | 1650 | *N/A* | *N/A* |
| | 403 | 100 | 1 | 6650 | 650.810 | *N/A* | 1 | 6650 | 3.034 | 1.547 |
| | | 120 | 20 | 5658 | 834.096 | 1349.540 | 40 | 5658 | 1.646 | 1.416 |
| | | 150 | 1 | 4454 | 6536.130 | *N/A* | 2 | 4454 | 2.510 | 1.171 |
| | | 200 | 1 | 3350 | 2252.040 | *N/A* | 1 | 3350 | 2.164 | 0.896 |
| ANI | 201 | 100 | 2 | 3300 | 369.090 | 2.828 | 2 | 3300 | 0.559 | 0.273 |
| | | 120 | 47 | 2771 | 342.869 | 1465.554 | 48 | 2771 | 0.238 | 0.268 |
| | | 150 | 50 | 2200 | 4.837 | 2.031 | 50 | 2200 | *N/A* | *N/A* |
| | | 200 | 50 | 1650 | 6.092 | 3.102 | 50 | 1650 | *N/A* | *N/A* |
| | 402 | 100 | 2 | 6650 | 1154.320 | 721.447 | 2 | 6650 | 3.022 | 1.481 |
| | | 120 | 20 | 5658 | 949.514 | 2124.473 | 40 | 5658 | 1.568 | 1.264 |
| | | 150 | 2 | 4454 | 10120.310 | 4907.519 | 2 | 4454 | 2.471 | 1.237 |
| | | 200 | 0 | 3350 | *N/A* | *N/A* | 0 | 3350 | 2.058 | 0.927 |

[1] Best integer lower bounds obtained by CPLEX after a 5h time limit ($L_{CPLEX}$) were used.

[2] Either $L_{CPLEX}$ or $L_3$ (Martello & Toth, 1990) lower bounds were used.

[3] Mean and standard deviation times for CPLEX are given only when optimal solutions could be reached.

[4] Mean and standard deviation times for the VNS heuristic are calculated for ten different seeds.

Table 2. Results for the tested instances

In Table 2, each line refers to the aggregate results given by a set of 50 instances. The results show that even for the smaller instances we could not obtain all the optimal solutions, especially for the capacity factor of 100%, which is not relaxed. However, the VNS heuristic could reach optimal solutions in much shorter runtimes using a regular PC solely, making it adequate to real applications faced by practitioners. Non-available (N/A) values presented in the VNS columns refer to instances that the initial solutions provided by the FFCD greedy heuristic were already optimal.

It should be noted that some caution should be exercised when attempting a direct comparison of running times between CPLEX and the VNS heuristic since CPLEX was run in parallel mode in a high-performance computing environment.

## 6. Conclusions

In this paper, the Bin Packing Problem with Compatible Categories (BPCC) was addressed, a variant of the BPP in which items have compatibility constraints and are divided into categories. The problem is common in last-mile distribution to nanostores located in large urban centers in developing countries.

To solve the BPCC, we proposed an algorithm based on the Variable Neighborhood Search (VNS) metaheuristic. Our VNS is simple to implement and allowed us to obtain the optimal solutions in very short CPU times for instances whose optimal results could be only obtained using an HPC that required substantial effort.

Further research should address larger instances of the problem, different compatibility matrices or unbalanced categories, to assess the VNS efficiency further and test it in a more comprehensive set of scenarios. In addition, comparisons with other heuristics and solution approaches might also be desired, as well as the development of different lower bound techniques focused on the BPCC and its characteristics.

**Acknowledgments**

The authors acknowledge the use of computational facilities of LCCA-Laboratory of Advanced Scientific Computation of the University of São Paulo (LCCA-USP), and the CNPq (Brazil's National Council for Scientific and Technological Development) financial support [grant numbers 830616/1999-3 and 304069/2010-8, respectively].